# Learning and teaching biological data science in the Bioconductor community

Jenny Drnevich[1,*], Frederick J Tan[2,*], Fabricio Almeida-Silva[3,4], Robert Castelo[5], Aedin C Culhane[6], Sean Davis[7], Maria A Doyle[6], Ludwig Geistlinger[8], Andrew R Ghazi[8], Susan Holmes[9], Leo Lahti[10], Alexandru Mahmoud[11], Kozo Nishida[12], Marcel Ramos[13], Kevin Rue-Albrecht[14], David JH Shih[15], Laurent Gatto[16], Charlotte Soneson[17,18]

[1]Roy J. Carver Biotechnology Center, University of Illinois Urbana-Champaign, Illinois, USA
[2]Johns Hopkins University, Department of Biology, Baltimore, Maryland, USA
[3]Department of Plant Biotechnology and Bioinformatics, Ghent University, Ghent, Belgium
[4]VIB Center for Plant Systems Biology, Ghent, Belgium
[5]Department of Medicine and Life Sciences, Universitat Pompeu Fabra, Barcelona, Spain
[6]Limerick Digital Cancer Research Centre, School of Medicine, University of Limerick, Ireland
[7]University of Colorado Anschutz School of Medicine, Denver, Colorado, USA
[8]Core for Computational Biomedicine, Department of Biomedical Informatics, Harvard Medical School, Boston, Massachusetts, USA
[9]Statistics Department, Stanford University, Stanford, California, USA
[10]Department of Computing, University of Turku, Finland
[11]Channing Division of Network Medicine, Harvard Medical School, Boston, Massachusetts, USA
[12]RIKEN Center for Biosystems Dynamics Research, 6-7-1 Minatojima Minamimachi, Chuo-ku, Kobe, Hyogo, Japan
[13]Department of Epidemiology and Biostatistics, City University of New York School of Public Health, New York, New York, USA
[14]MRC WIMM Centre for Computational Biology, MRC Weatherall Institute of Molecular Medicine, University of Oxford, Oxford, UK
[15]School of Biomedical Sciences, Li Ka Shing Faculty of Medicine, The University of Hong Kong, Hong Kong SAR, China
[16]Computational Biology and Bioinformatics Unit, de Duve Institute, UCLouvain, Brussels, Belgium
[17]Friedrich Miescher Institute for Biomedical Research, Basel, Switzerland
[18]SIB Swiss Institute of Bioinformatics, Basel, Switzerland
*These authors contributed equally
Correspondence to laurent.gatto@uclouvain.be (LG) or charlotte.soneson@fmi.ch (CS)

## Abstract

Modern biological research is increasingly data-intensive, leading to a growing demand for effective training in biological data science. In this article, we provide an overview of key resources and best practices available within the Bioconductor project - an open-source software community focused on omics data analysis. This guide serves as a valuable reference for both learners and educators in the field.

# Introduction

Modern biological research relies heavily on high-throughput technologies, including sequencing, imaging, cytometry and mass spectrometry. These technologies generate vast amounts of data that require multi-disciplinary teams and sophisticated computational methods for analysis and interpretation (1). To meet the increasing demand for well-trained data scientists in biology, substantial efforts are being directed towards establishing and disseminating pedagogical best practices (2–4).

Bioconductor, established in 2001, is a powerful and widely used open-source software community project for biological data analysis (5–7). It offers a comprehensive collection of over 2,300 R packages with specialized data structures and analysis methods for biological data. Bioconductor ensures reliability and robustness of its packages via an automated build system that runs daily checks on all packages for code quality, documentation completeness, and adherence to Bioconductor standards. This makes it attractive to developers, thousands of whom have already contributed R packages to the project. Its comprehensive suite of data analysis methods makes it valuable to researchers and annual download estimates exceed 1 million (8,9).

In light of Bioconductor's broad scope, navigating its extensive ecosystem to locate and effectively utilize desired packages can be challenging for researchers. To foster efficient discovery and use of its resources, the Bioconductor Training Committee was established in 2020 to streamline and coordinate educational initiatives (10). The goals of the committee include providing a meeting place for community members interested in training, advocating for maintenance of important existing material, identifying gaps in current material, and coordinating training activities, including with other bioinformatics communities.

The best answer to the question "How do I get started with Bioconductor?" depends on the person's goals and background. This manuscript provides an overview of the many resources that the Bioconductor community has developed. The first part takes the perspective of a learner and suggests suitable entry points depending on the aim. The second part instead takes the perspective of an instructor and outlines the resources and community available to assist with delivering Bioconductor-related training.

# Learning Bioconductor

One of the foundational ideas of R, and later of Bioconductor, was to facilitate the passage from user to developer (11). This requires a healthy and coordinated community, as well as effective learning material.

## Prerequisites

Using Bioconductor packages requires basic knowledge of the R language. From a data analysis perspective, learners need familiarity with tabular data, basic understanding of exploratory data analysis, and comfort with sequential scripting (taking the output of one command and using it as input for the next command). Some of these skills may be acquired through formal coursework or other initiatives like The Carpentries (12), R for Data Science (13), swirl (14), or Massive Open Online Courses (MOOCs) (15,16). As detailed below, the Bioconductor Training Committee has developed an "Introduction to data analysis with R and Bioconductor" (*bioc-intro*) workshop addressing basic R concepts from a genomic data standpoint (17).

## Acquire the Fundamentals

The R and Bioconductor community produces learning resources for various target audiences (Table 1). For novice R users, the "Introduction to data analysis with R and Bioconductor" (*bioc-intro*) workshop provides a great entry point (17). Domain-specific introductory material also includes DFCI YES for CURE, which is intended for use by secondary and undergraduate students undertaking cancer data science studies (18). For experienced R users, we recommend familiarizing with core Bioconductor data containers, such as GenomicRanges (19) or SummarizedExperiment (5), to use other Bioconductor packages effectively. We refer the reader to Table 1 and the remainder of the manuscript to pinpoint which resources would be most beneficial based on individual learning goals.

Bioconductor community members also develop and organize numerous *workshops and courses* every year, many of which are listed (together with links to the material) on the Bioconductor website (20). For example, for users with a good knowledge of R who would like to learn the basics of genomics data analysis, longer workshops and summer schools are organized yearly (e.g., (21,22)).

**Table 1. Notable resources for learning and teaching Bioconductor**

| Goal | Acquire the Fundamentals (novice R) | Acquire the Fundamentals (experienced R) | Analyze Your Data | Connect with the Community | Train Others | Develop a Package |
|---|---|---|---|---|---|---|
| Bioconductor Carpentry | | | | | | |
| *bioc-intro* | X | | | X | X | |
| Topic specific (e.g. *bioc-rnaseq, bioc-scrnaseq*) | | | X | X | X | |
| *bioc-project* | | X | | X | X | X |
| Bioconductor Conferences | | | | | | |
| Keynotes/Talks | | | X | X | | X |
| Package Demos/Workshops | X | X | X | X | X | X |
| Networking | | | | X | | |
| Other Workshops | | | | | | |
| Longer courses (e.g. CSAMA, CSHL) | | X | X | X | | |
| Local workshops, Galaxy Training Network, commercial offerings, etc | X | X | X | X | | |
| Packages | | | | | | |
| Help pages | | X | X | | | |
| Vignettes | X | X | X | | | |
| Workflows | | X | X | | | |
| Datasets | | X | | | X | X |
| GitHub Issues | | | X | | | X |
| Books | | | | | | |
| R for Data Science | X | | | | X | |
| Bioconductor Books | | | X | | X | |

| Goal | Acquire the Fundamentals (novice R) | Acquire the Fundamentals (experienced R) | Analyze Your Data | Connect with the Community | Train Others | Develop a Package |
|---|---|---|---|---|---|---|
| R packages/Advanced R | | X | | | | X |
| Bioconductor contributor guide | | | | | | X |
| Community | | | | | | |
| Support Site | X | X | X | X | X | |
| Slack | X | X | X | X | X | X |
| Developer Mailing List | | | | X | | X |
| Bioconductor YouTube channel | | | X | | X | |
| Special Interest Groups (e.g. R-Ladies, R User Groups) | | | | X | | |

## Analyze Your Data

Locating the right package and method for a specific data analysis task can be difficult. A large number of packages and functions are available, and often several packages provide overlapping functionality that may be equally suitable for the task at hand. One entry point for finding packages suitable for a specific type of analysis are the biocViews (23). Using these, it is possible to filter the list of packages based on keywords related to, for example, the type of data, biological question, or statistical approach at hand. To further improve resource discoverability, Bioconductor is actively exploring the integration of biocViews with EDAM (a controlled vocabulary for bioinformatics concepts) (24) and enhancing the search functionality on the Bioconductor website and its documentation. These efforts aim to help users more efficiently locate relevant packages and workflows.

Once suitable packages have been found, Bioconductor provides multiple levels of documentation. Each function must have a *manual* page, documenting what it does, its inputs and outputs, and often executable examples with additional guidance and references. Each package must also have at least one *vignette* showing how to use it to run a typical analysis. This enables a "learning by doing" philosophy, where users can repeat and adjust the code provided in examples and vignettes, using either their own data or the example data used by the package developer. Video recordings of many package presentations are available via Bioconductor's YouTube channel (25).

Bioconductor also features *workflows* (26), dedicated to end-to-end analysis of specific types of data, typically using many different software packages. While these are often used as self-study material, they also form the basis for instructor-led courses, and some have been published in online journals (e.g. (27,28)). Finally, several *books* have been written about Bioconductor (29). Early books (30) were targeted towards analysis of microarrays and the new challenges that arose for researchers confronted with the complexity of heterogeneous data. More recently, various online books including ones discussing analysis of single-cell data (31), spatial transcriptomics data (32), Hi-C data (33) and microbiome data (34), as well as a broader treatise on modern statistics for modern biology (35) have been added to the collection. Compared to the workflows, the books cover a broader scope and contain more discussions about concepts. Both workflows and online books distributed via Bioconductor are regularly built and tested, and thus a user can be assured that the code within them is executable, a feat difficult to achieve with statically provided teaching material that often gets out of date as packages are updated.

The development of these long-form narrative documentation formats has greatly benefited from the development of general publishing tools within the R ecosystem, including R Markdown (36), bookdown (37), pkgdown (38) and Quarto (39), all enabling the application of literate programming (40) to generate fully reproducible, human-readable examples and documentation.

Finally, users may want to interact with other popular analysis ecosystems, and Bioconductor provides tools and resources to facilitate interoperability. For example, the tidyomics ecosystem (41) was developed to bridge Bioconductor with the popular R tidy programming paradigm (42). Additionally, many tools for the analysis of single-cell and spatial omics data, such as the ones provided by the scverse consortium (43), are written in Python. To facilitate interoperability between Python and R, basilisk (44) simplifies the management of self-contained conda environments for Python packages, and zellkonverter (45) enables the interconversion between Bioconductor SingleCellExperiment objects and Python anndata objects (46). The Bioconductor Carpentry single cell module (47) provides hands-on guidance for interoperability with Seurat (48) and Scanpy (49). The first SpatialData hackathon and workshop organized by the scverse consortium in November 2024 brought together R/Bioconductor developers and Python experts to enhance interoperability and scalability of spatial omics frameworks, highlighting the collaborative efforts between the Bioconductor and scverse communities (50). Furthermore, the 2025 Galaxy and Bioconductor Community Conference (GBCC 2025) will focus on advancing bioinformatics and data science tools across diverse platforms, promoting interoperability and collaboration between the Galaxy and Bioconductor communities (51).

## Connect with the Community

In day-to-day data analysis practice, questions commonly arise. These questions are sometimes technical (e.g., related to the precise execution of a specific function) and sometimes more conceptual (related to the interpretation of results or best practices for specific tasks). In both cases, learning from other's experiences can be immensely helpful. The Bioconductor community provides multiple venues for such interactions. The *support forum* (52) is aimed at questions related to the use of Bioconductor packages, while the *developer mailing list* (53) is focused on package development-oriented questions. In addition, the *Slack workspace* (54) offers channels dedicated to a wide variety of topics. Individual packages are often developed on GitHub, where users can raise issues to report bugs or request features. In all cases, discussions are happening in public - hence, other community members with similar questions can also benefit, and over time an extensive knowledge bank is collectively built up.

Other avenues to connect with community members and learn about Bioconductor are the yearly *conferences*, which gather developers and users to discuss upcoming developments and listen to keynote presentations, contributed talks and package demos. A yearly conference has been held in North America since 2005 and regional conferences were established for Europe in 2007 and Asia in 2015.

## Develop a Package

For many community members, their entry point into the Bioconductor project comes from the desire to learn how to use the provided packages for analyzing data. Others get into the project by contributing a package, often implementing a statistical method or computational pipeline they established. Developing a Bioconductor package requires different skills than using the packages, and thus different teaching material.

The R community provides several excellent resources for package development, which are equally relevant for Bioconductor packages (55,56). Bioconductor also provides extensive guidelines on developing and maintaining packages, as well as coding style and how to interact with other languages like C and Python (57,58). Bioconductor strongly encourages the re-use of community-established object classes for representation of data to maximize interoperability between packages, optimize computational efficiency, facilitate user learning, and reduce maintenance efforts for developers. The Carpentries-style "Introduction to the Bioconductor project" lesson (*bioc-project*) provides an overview of common object classes in Bioconductor such as SummarizedExperiment and GenomicRanges (59).

Upon submitting their package to Bioconductor, developers are assigned a reviewer - an experienced package developer - who provides a thorough assessment and constructive feedback to enhance the package. In 2020, Bioconductor launched a community mentorship program (60), offering first-time developers regular, tailored advice from a more experienced developer. Community members with package development experience can apply to become part of the team of reviewers. The Bioconductor contributor's guide describes the procedure, expectations and onboarding process, and provides additional helpful resources for package reviewers (61).

## Teaching Bioconductor

The Bioconductor Training Committee was established in early 2020 to coordinate education-related activities, form a community of instructors trained in teaching best practices, survey the available training material, and develop strategies to fill identified gaps. The committee is open to any interested community member and holds monthly virtual meetings where topics related to Bioconductor and training are discussed. To gather further feedback, a set of community calls were organized during a "Teaching Week" in 2022.

### Carpentries Global Instructor Training Program

A close interaction with The Carpentries was initiated early on, and the first five community members were certified as Carpentries Instructors in 2020. Supported by a grant from The Chan Zuckerberg Initiative (CZI), in August 2022, Bioconductor officially joined The Carpentries as a member organization, enabling the certification of more community members (62) (Fig 1A). The grant aimed to fund the training of 30 Instructors, distributed around the world, and also hiring a Community Manager to coordinate the program. During the first year, 18 Instructors were trained, and after the second year, 31 Instructors have now fully completed their training and earned certification (63); up-to-date Instructor information is tracked in a GitHub repository, and includes additional interested community members who received their certification independently (64).

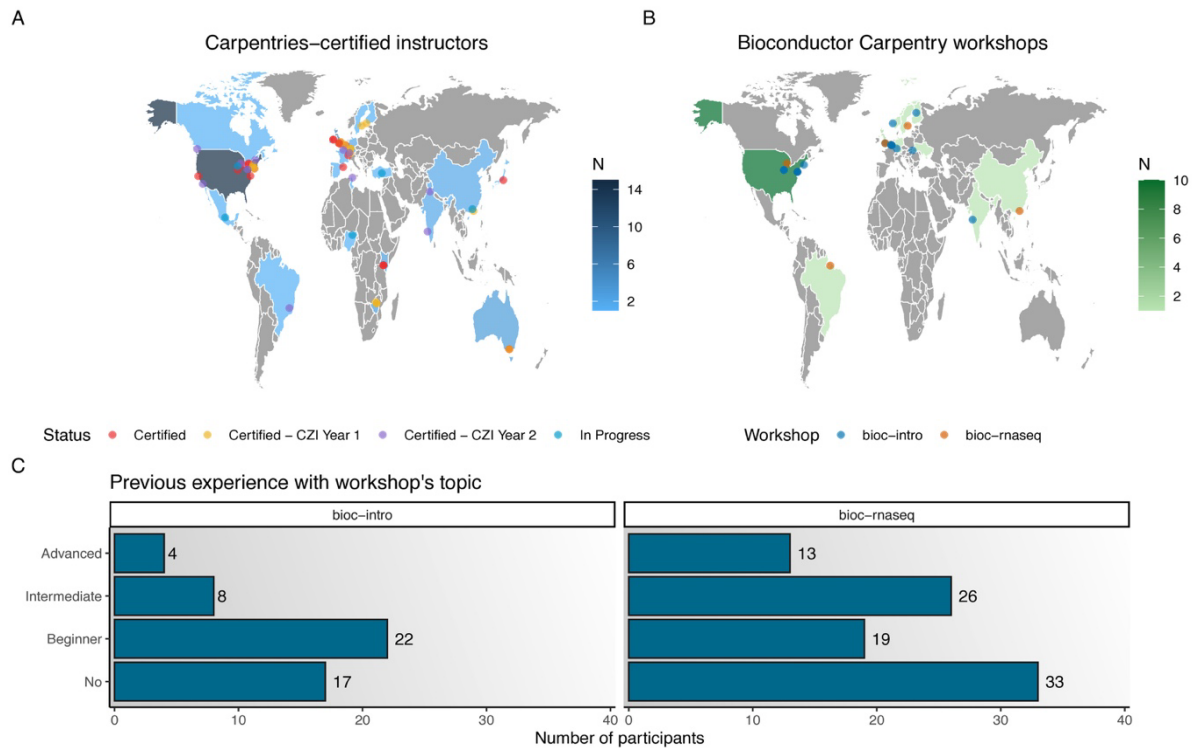

**Fig 1. Geographic distribution of Carpentries-certified Instructors and delivered workshops, as well as learner profiles. A.** Geographic distribution of Bioconductor Carpentries-certified Instructors. Color intensities indicate the total number of Instructors (certified, certified owing to CZI grant, and certification in progress) in each country. **B.** Geographic distribution of workshops taught using the Bioconductor Carpentry material. **C.** Previous experience of participants with Bioconductor Carpentry workshop's content at pre-conference workshops (*bioc-intro*: 51 participants over 3 offerings, *bioc-rnaseq*: 111 participants over 5 offerings). Topic-specific workshops appear to draw more participants with intermediate or advanced skills. The base layer of the maps in panels **A** and **B** was obtained from Natural Earth v2.0.0 (65), via the maps R package v3.4.2 (66).

Thanks to further support from CZI, we are continuing to expand our Bioconductor Carpentry training program, aiming to build local capacity and address specific training needs in underserved regions, with an initial focus on Africa. Our goal is to ensure equitable and accessible workshops, fostering a vibrant global Bioconductor community. For more details, please refer to our blog post (67) and the Nairobi workshop course page (68).

## Bioconductor Carpentry Curricula

In its first year, the Training Committee initiated the development of three lessons within the Carpentries Incubator framework (69). The first (*bioc-intro*) provides an introduction to data analysis with R, geared towards analyzing high-throughput biological data with Bioconductor (17). It is based upon material developed for a previous Data Carpentry lesson on data analysis with R in ecology, but the data set as well as certain parts of the content have been modified to better align with our purposes. The second lesson (*bioc-project*) summarizes the various components of the Bioconductor project, and includes episodes on getting help and navigating the documentation, installing packages, and introductions to some of the important object classes (59). The third lesson (*bioc-rnaseq*) covers bulk RNA-seq data analysis using Bioconductor (70). This lesson assumes basic familiarity with R and an understanding of the motivations and technologies involved in RNA-seq experiments. A fourth lesson (*bioc-scrnaseq*), covering single-cell RNA-seq analysis, has since been added (47). This lesson demonstrates how to use Bioconductor tools for essential single-cell analysis steps with a focus on working with large data, interoperability with other popular analysis ecosystems, and accessing public data as, e.g., available from the Human Cell Atlas.

The Bioconductor Carpentry lessons so far have been taught at least 20 times as pre-conference workshops at the three yearly Bioconductor conferences, and at separate events, since 2022 (Fig 1B), reaching over 300 participants. Expectedly, most participants have no or little experience with the workshop content (i.e., R or RNA-seq data analysis), but all workshops had participants with intermediate or advanced skills (Fig 1C). Participants with intermediate or advanced skills were either self-taught learners who wanted to (re-)learn from experts to gain confidence and get to know best practices, or experienced researchers who can analyze data, but claimed to lack knowledge of theoretical details (e.g., normalization techniques, statistical models in differential expression analyses, more complex design matrices, proper handling of confounders, etc).

## Computational Infrastructure

Bioconductor aims to ensure that any aspiring educator or learner can contribute to or learn from the collection of educational workshops. Unfortunately, disparities in access to computers and the need for specific software versions can become a learning barrier for some, disproportionately affecting those from low-resourced institutions, regions, and/or backgrounds.

We aimed to solve this problem by delivering workshops as pre-configured containerized environments and making them widely available as a service. In 2023, Bioconductor began hosting a Workshop Service (71) as a free platform accessible for any member of the community to run through workshop material contributed by fellow community members. This service is a modified instance of the popular Galaxy software (72), and allows users to launch a private pre-configured RStudio instance for any of the contributed workshops at no cost to users or the project. Attending workshops is thus made accessible to anyone with a web browser, as our Workshop Service allows them to run code in a pre-configured, predictable, and reproducible environment regardless of the type of device or computational power available to each user.

For workshop creators, Continuous Delivery (CD) automation is used to ease the path for adding to the workshop collection. New workshops can be added in real time with no service disruption to active users. The workshop list is source-controlled in a public GitHub repository, and automation is provided for anyone to directly contribute workshops. Workshop instructors are recommended to use a modular template GitHub repository (BuildABiocWorkshop (73)) to build and distribute workshop instances. The repository contains a set of GitHub Actions for building, testing, and creating workshop container images for distribution. It provides instructors with increased flexibility to install the necessary system and R package dependencies for their workshops.

## Translations

Most educational material and documentation distributed via the Bioconductor project is written in English. However, this limits the accessibility, especially among user communities where English is not a common language. So far, the only content of Bioconductor systematically translated into multiple languages has been its project-wide code of conduct (74).

To initiate the translation of the community-developed lesson material into other languages, we set up Crowdin, a localization management platform, for our organization (75). Crowdin is a cloud-based tool that allows contributors from all over the world to work together to translate documents into different languages. The translation can be done in a web browser, where contributors can focus on the translation without worrying about technical details. We have also set up automatic machine translation (using DeepL (76) and Google translation) for the three initial Bioconductor Carpentry lessons, allowing us to draft initial translations for later review by native speakers.

## Using AI/LLMs in teaching

Generative AI and large language models such as ChatGPT open interesting avenues for teaching and learning. We note that this is a very active area of research where the training community is still working to understand how to best use this for teaching (77). As with all tools, understanding their proper use and how to formulate effective queries and interpret results correctly is essential. Of note, this emphasizes the need to learn correct and precise terminology. On the other hand, in classroom teaching excessive use of jargon is often avoided in favor of more conceptual explanations to reduce cognitive overload. However, the new power of large language models can be used as a motivator for learning such jargon and buzzwords because the use of these words in prompts greatly enhances the quality of the responses and thus students can observe the effectiveness of this precision in language usage in real time.

The implementation of specifically-trained Bioconductor chatbots holds enormous potential for democratizing access to coding and analysis support within the Bioconductor community and will be empowering for new users that might be hesitant to post questions on a public online forum. Currently available generalist models, such as Claude and ChatGPT, have already shown impressive capabilities in generating R code and, for example, assisting with the creation of R/Shiny apps in Posit's ShinyAssistant (78). Specifically, writing instructive comments at the beginning of a code section typically improves the quality of the suggestions made by systems such as Copilot or Claude when using these tools within the coding environment. Augmentation techniques such as prompt engineering, fine-tuning, and retrieval-augmented generation (RAG) can be employed for the creation of grounded models that improve response accuracy over generalist models and that are less prone to hallucinations. An important task is the benchmarking of the different models on curated evaluation datasets. The Bioconductor support site (52) records user questions and expert answers on all aspects of the project for more than two decades, providing potential ground truth for the evaluation of the quality of answers generated by the different models.

## Discussion

The growing volume and complexity of biological data sets, coupled with the diverse backgrounds of researchers tasked with their analysis, highlight the need for high-quality and accessible learning material. For example, many modern biomedical projects involve integration of multiple high-throughput molecular data types, requiring training material focusing specifically on such integrative tasks. Increases in data volume also require users to work in new environments, such as servers or high-performance computing (HPC) environments, or use disk-backed data structures. Additionally, some parts of the analysis may require the use of a different programming language than R, such as Python or C++.

A key challenge that the Bioconductor Training Committee aims to address is organizing decentralized efforts to develop new and maintain existing documentation and teaching content. As tasks get increasingly complex and new technologies emerge, the number of instructors with adequate expertise to compile and deliver corresponding training material decreases, which can lead to a large, unmet training need. This can be mitigated by making training material FAIR (findable, accessible, interoperable and reusable). To this end, a working group (79) has been established within Bioconductor to improve discoverability and accessibility of training material via TeSS (80). To motivate and compensate decentralized efforts, we are partnering with other open-source projects like Galaxy (81) to organize sprints like the Galaxy Smörgåsbord and continually seek funding from organizations like CZI for additional Carpentries Instructor certifications. Strategies to monitor content maintenance include explicit tagging of maintainers and continuous integration and automated testing of individual package vignettes, though we note that workflows and books depending on a large number of packages remain a challenge to maintain, especially since funding opportunities for such efforts are scarce. Another avenue that is being explored by the training committee is the consolidation and creation of short "How To" documents, each illustrating how to use Bioconductor packages to solve a very specific question. These documents may be useful to users who have a concrete problem to solve but are unsure about which packages would be most suitable to use.

A major lesson learned during the formation of the Training Committee was the benefits of partnering with The Carpentries, an organization already dedicated to the pedagogy of how to best teach computational skills. They provided support and guidance in preparing effective workshop material, including periodic assessments, and how best to use the material in teaching others. Beyond the practical, their instructor training lessons are immensely powerful to help improve not only novice but also experienced instructors. This partnership has also empowered instructors to independently organize Bioconductor Carpentry workshops, with events taking place in Latin America, North Africa, Asia, Europe, and the USA. By drawing on shared resources and experiences, these efforts have fostered a collaborative and supportive training community. Collaborating with The Carpentries ensured Bioconductor could provide high-quality, effective training while accelerating the development of its training initiatives.

We believe that Bioconductor, with a diverse community of expert developers and users, and a strong technical infrastructure, is well-positioned to deal with the challenges outlined above. The Training Committee is well suited to help organize and coordinate training efforts and development and maintenance of educational material and ensure that it is disseminated widely to be useful to as many scientists as possible. Finally, we encourage new members to engage with the Bioconductor community. Whether you are interested in contributing to our projects, attending workshops, or participating in discussions, there are numerous ways to get involved. Join our meetings, engage with us on Slack, or attend our upcoming workshops. We welcome your participation in our efforts to build a vibrant and inclusive Bioconductor community.

## Acknowledgements

The authors would like to thank all the current and former members of the Bioconductor Training Committee for their valuable contributions. We thank Vincent Carey for feedback on the manuscript, and François Michonneau and Toby Hodges from The Carpentries for invaluable guidance as we developed the Bioconductor Carpentry program. This project has been made possible in part by grants 2021-237919 (to AC), 2022-311145 (to RC) and 2024-342820 (to AC) from the Chan Zuckerberg Initiative DAF, an advised fund of Silicon Valley Community Foundation. LL acknowledges funding from the Research Council of Finland (decision 330887) and the European Union's Horizon 2020 research and innovation programme under grant agreement No 952914. SD acknowledges funding from NCI grant 1U24CA289073. AM acknowledges funding from NIH grant 2U24HG004059-17. CS is supported by the Novartis Research Foundation. The funders had no role in study design, data collection and analysis, decision to publish, or preparation of the manuscript.

## Competing Interests

The authors have declared that no competing interests exist.

## Data Availability

Data and code for Fig 1 are available online at https://doi.org/10.5281/zenodo.14995886. All other data are in the manuscript and in the Supporting Information.

# Supporting Information

### S1 Table. URLs for Resources Mentioned in the Manuscript

| Introduction | |
|---|---|
| Bioconductor Website | https://www.bioconductor.org |
| Bioconductor Training Committee | https://training.bioconductor.org |
| | |
| **Prerequisites** | |
| The Carpentries | https://carpentries.org |
| R for Data Science | https://r4ds.hadley.nz |
| swirl: Learn R, in R | https://swirlstats.com |
| Data Science Specialization | https://www.coursera.org/specializations/jhu-data-science |
| Data Analysis for the Life Sciences | https://www.edx.org/xseries/data-analysis-life-sciences |
| | |
| **Acquire the Fundamentals** | |
| Courses & Conferences | https://bioconductor.org/help/course-materials |
| Biological Data Science | https://csama2024.bioconductor.eu |
| Statistical Analysis of Genome Scale Data | https://meetings.cshl.edu/courses.aspx?course=C-DATA |
| Yes for CURE | https://vjcitn.github.io/YESCDS/ |
| | |
| **Analyze Your Data** | |
| YouTube Channel | https://www.youtube.com/user/bioconductor |
| Workflows | https://www.bioconductor.org/packages/release/workflows |
| Books | https://bioconductor.org/help/bioconductor-books |
| Single-Cell Analysis | https://bioconductor.org/books/release/OSCA |
| Spatial Transcriptomics Analysis | https://lmweber.org/BestPracticesST |
| Hi-C Analysis | https://bioconductor.org/books/release/OHCA |
| Microbiome Analysis | https://microbiome.github.io/OMA |
| Modern Statistics for Modern Biology | https://www.huber.embl.de/msmb/ |
| | |
| **Connect with the Community** | |
| Support Forum | https://support.bioconductor.org |
| Developer Mailing List | https://stat.ethz.ch/mailman/listinfo/bioc-devel |
| Slack Workspace | https://slack.bioconductor.org |
| | |
| **Develop a Package** | |
| Contributor Guide | https://contributions.bioconductor.org |
| Developer Mentorship Program | https://www.bioconductor.org/developers/new-developer-program |
| Advanced R | https://adv-r.hadley.nz/ |
| R Packages | https://r-pkgs.org/ |
| | |
| **Carpentries Global Instructor Training Program** | |
| Announcement | https://blog.bioconductor.org/posts/2022-07-12-carpentries-membership |